\documentclass[prd,twocolumn]{revtex4}
\usepackage{graphicx}
\usepackage{amssymb}
\usepackage{epstopdf}
\usepackage{hyperref}

\begin{document}

\title{Neutron Stars in $f(R)$ Gravity with Perturbative Constraints}
\date{\today}

\author{Alan Cooney}
\affiliation{Department of Physics, University of Arizona, 
Tucson, AZ 85721, USA}

\author{Simon DeDeo}
\affiliation{ Santa Fe Institute, 1399 Hyde Park Road, Santa Fe, NM 87501, USA}

\author{Dimitrios Psaltis}
\affiliation{Departments of Astronomy and Physics, 
University of Arizona, Tucson, AZ 85721, USA}

\begin{abstract}
We study the structure of neutron stars in $f(R)$ gravity theories
with perturbative constraints. We derive the modified
Tolman-Oppenheimer-Volkov equations and solve them for a polytropic
equation of state. We investigate the resulting modifications to the masses 
and radii of neutron stars and show that observations of surface phenomena
alone cannot break the degeneracy between altering the theory of gravity versus
choosing a different equation of state of neutron-star matter. On the other hand, 
observations of neutron-star cooling, which depends on the density of matter 
at the stellar interior, can place significant constraints on the parameters of the theory.
\end{abstract}

\maketitle 

\section{Introduction}
\label{INTROsec}
Recent interest in modified theories of gravity has been spurred by
the discovery that the Universe is undergoing accelerated expansion
(see, e.g.,~\cite{P99,R98,WMAP09}). The simplest solution consistent 
with these observations
posits a cosmological constant $\Lambda$. The magnitude of this
cosmological constant is significantly less than what was expected,
and many undertakings have been made to see if there are plausible
alternative explanations~\cite{CDTT04,S08}. Outstanding questions also
present themselves in the formation of
singularities~\cite{P08} and the seeming contradiction between quantum
mechanics and gravity in the context of black hole
thermodynamics~\cite{W01}. All these suggest that there may yet be
much to understand about the nature of gravity at extreme-curvature
scales, far removed from our everyday experience.

The two most popular approaches to modifying gravity have been the
introduction of an additional scalar field (e.g.~\cite{PR03}), or the
related approach of replacing the Einstein-Hilbert action with a
general function of the Ricci scalar $f(R)$
(e.g.~\cite{CDTT04}). Within either framework the additional scalar
degree of freedom can be tuned to mimic the cosmological constant,
or any type of cosmological evolution at cosmological
scales~\cite{W07}.

Despite the premise of such modifications, the non-linear character of
gravitational theories has proven a significant obstacle to
introducing new dynamical fields to drive modifications to gravity at
the cosmological scale without the same fields reemerging at widely
different curvature scales. One such example is the problem of
ensuring that $f(R)=R\pm\mu^4/R$ theories pass the current Parametrized
Post-Newtonian (PPN) bounds. When the new field is dynamical, the PPN
parameter $\gamma$ is forced to a value of $1/2$, which is very far
from the present experimental bound~\cite{C03}. As a result one has
to choose a function $f(R)$ only from the class which can
adequately suppress the new dynamical field on solar-system
scales. The chameleon mechanism~\cite{KW04,FTB07,HS08} provides such an
alternative.

In addition to the PPN constraints, instabilities related to the
functional form of $f(R)$ have also been studied at length. This is
especially true for the Dolgov-Kawasaki instability~\cite{DK03},
which requires that $\partial ^2 f/ \partial R^2 > 0$ in order that the
effective mass of the equivalent scalar degree of freedom be positive.
In the strong-field regime, recent results~\cite{KM08} suggest that
this very choice may well prohibit the formation of compact objects
above a curvature scale readily observed.  However, the fatal curvature
singularity may be avoided by the chameleon mechanism~\cite{BL09,UH09}.

Perhaps the source of the instabilities and consistency issues
many of these models encounter is the result of treating these
modifications as though they are exact. The original motivation behind
introducing additional functions of the curvature was to generate a
new phenomenology at a specific scale. However, many of the problems
encountered by $f(R)$ gravity theories originate at curvature scales
far removed from the ones under consideration.  An alternative
formulation for handling corrections to General Relativity is to view
the new terms as only the next to leading order terms in a larger
expansion. In this context there is no reason to suspect that the new
phenomenology is due to new dynamical fields. The technique for
handling a field expansion of this form is well developed~\cite{EW89}
and is known as perturbative constraints or order
reduction~\cite{JLM86}.

Gravity with perturbative constraints allows us to explore alternative
phenomenologies of gravity while maintaining important consistency
conditions including gauge invariance, the assumption that we are approximating a fundamentally second order field theory, and the conservation of stress-energy. Maintaining such constraints while enlarging the space of possible behaviors of gravitation is the goal also of the Parametrized Post-Friedman approach~\cite{HS07,H08,FS10}.

In previous works~\cite{DP08, CDP09}, we have analyzed the effect
of treating $f(R)$ models of gravity via perturbative constraints
primarily at cosmological scales. In this paper, we examine the
ramifications of modifications to gravity in the context of compact
objects. We show how the method of perturbative constraints allows for a consistent phenomenology for gravity on both large (Hubble-length perturbations linear in metric variables, but strongly relativistic, $L\sim c/H_0$) and small scales (stellar scales, non-linear in metric perturbations, and strongly relativistic, $GM/rc^2\sim 1$.)

The layout of this work is as follows. In Section~\ref{PCsec}, we review
the equations of $f(R)$ gravity treated with perturbative
constraints. In Section~\ref{SPCsec}, we derive the modified
Tolman-Oppenheimer-Volkov equations and show that the exterior
solution is the Schwartzchild-de Sitter metric. In Section~\ref{MRsec}, we
demonstrate that such objects are stable and we solve numerically for
their mass-radius relation for a polytropic equations of state. Finally in 
Section~\ref{DIS} we discuss how we can discriminate modifications to 
gravity from uncertainty in the neutron star equation of state.

\section{Perturbative Constraints}
\label{PCsec}

Gravity with perturbative constraints~\cite{EW89} (or
order-reduction~\cite{JLM86}) is a technique for treating  equations of motion
that appear higher than second order, where the origin of the
higher derivatives can be traced to the truncation of an infinite
series expansion. Such a situation can arise with non-local theories 
as well as effective field theories.

In the context of $f(R)$ gravity theories, we parametrize the
deviation from General Relativity by a single parameter $\alpha$ and derive
the equation of motion from a covariant action
\begin{eqnarray}
 \label{action}
 {\it S} = \frac{1}{16\pi}\int d^4x&\sqrt{-g}&\left[R-2\Lambda + \alpha f(R)+\mathcal{O}(\alpha^2)\right] \nonumber \\
 &&+  {\it S_M}(g_{\mu\nu},\psi)\;,
\end{eqnarray}
with $G=c=1$. Here $g_{\mu \nu}$ is the metric, $g$ its determinant,
and $R$ the Ricci scalar.  We denote any additional terms above order
$\alpha$ by $\mathcal{O}(\alpha^2)$. We may not impose any
constraints at the level of the action without altering the nature of
the variational principle. The resulting field equation is
\begin{eqnarray}
\label{EOM1}
R_{\mu \nu}-\frac{1}{2}g_{\mu \nu}R +g_{\mu \nu} \Lambda&+ &\alpha \left[f_R R_{\mu \nu} - \frac{1}{2}g_{\mu \nu}f -\right. \nonumber \\
 \left. \left(\nabla_{\mu}\nabla_{\nu} -g_{\mu \nu}\square \right)\frac{}{}f_R \right]&+&\mathcal{O}(\alpha^2)= 8\pi T_{\mu \nu} \;,
\end{eqnarray}
where $f_R \equiv \partial f / \partial R$.

At zeroth order in $\alpha$, these equations are second order in the
metric; we denote the solution at this order by $g^{(0)}_{\mu
\nu}$. We then solve the system for the higher order terms by writing
\begin{equation}
\label{expansion}
g_{\mu \nu } = g^{(0)}_{\mu \nu}+\alpha \;g^{(1)}_{\mu \nu}+ \mathcal{O}(\alpha^2) \;.
\end{equation}
The perturbative consistency of this approach is guaranteed to order
$n$ provided $\alpha^{n+1} \;g^{(n+1)}_{\mu \nu} \ll g^{(0)}_{\mu \nu}+\dots + \alpha^{n}\;g^{(n)}_{\mu
\nu}$, as we outlined in a
previous paper~\cite{CDP09}. Note that this condition is not to be
understood as requiring the product $\alpha f(R)$ to be necessarily
smaller in magnitude than $R$.

For the purposes of this work it will prove useful to rewrite
Eq.~(\ref{EOM1}) using its trace
\begin{equation}
\label{TR}
R - \alpha \left[f_R R - 2f +3\square f_R \right]+\mathcal{O}(\alpha^2)= -8\pi T + 4 \Lambda \;.
\end{equation}
Substituting the Ricci scalar $R$ from the above equation into
Eq.~(\ref{EOM1}) gives
\begin{eqnarray}
\label{EOM2}
R_{\mu \nu}- g_{\mu \nu} \Lambda+ \alpha \left[f_R R_{\mu \nu} - \frac{1}{2}g_{\mu \nu}\left(f_R R-f\right) -\right. \nonumber \\
 \left. \left(\nabla_{\mu}\nabla_{\nu} +\frac{1}{2}g_{\mu \nu}\square \right)\frac{}{}f_R \right]+\mathcal{O}(\alpha^2)=8\pi\left(  T_{\mu \nu} -\frac{1}{2}g_{\mu \nu}T\right) \;.
\end{eqnarray}
This is the form of the field equation we will be using. Henceforth
we shall understand the equality sign to mean equality
up to order $\alpha$ and drop the explicit use of
$\mathcal{O}(\alpha^2)$.

\section{Stars with Perturbative Constraints}
\label{SPCsec}

The metric of a static, spherically symmetric object can always be
written in the form
\begin{equation}
\label{METRIC}
\small{ds^2 = -B(r)dt^2 + A(r)dr^2+r^2\left( d\theta^2 + \sin^2\theta d\phi^2\right)} \;,
\end{equation}
where $B(r) = B^{(0)}(r)+ \alpha B^{(1)}(r)+ \dots$, $A(r) =
A^{(0)}(r)+ \alpha A^{(1)}(r)+ \dots$, and $B^{(0)}(r)$ and
$A^{(0)}(r)$ are the general relativistic metric elements.

For the purpose of this paper we presume the form $f(R) \propto
R^{n+1}$ for an integer $n \ne 0,-1$. We shall also assume that the
 energy-momentum tensor within the star is
that of a perfect fluid.  Following our previous
studies~\cite{CDP09} we find it convenient to express the
$\mathcal{O}(\alpha)$ correction in terms of the derivative $f_R$. The
first three field equations are
\begin{eqnarray}
\label{00}
\frac{R_{00}}{B}&+& \alpha f_R\left\{ \frac{R_{00}}{B}+\frac{R}{2}\left(\frac{n}{n+1}\right) -\frac{n}{2A}\left[ -\frac{R''}{R}-n\frac{R'^2}{R^2}+\right.\right. \nonumber \\
\frac{R'^2}{R^2}&+&\left.\left.\frac{R'}{R}\left(\frac{A'}{2A}-\frac{3B'}{2B}-\frac{2}{r}\right)\right]\right\} = 4\pi\left(\rho+3P\right)-\Lambda \;,
\end{eqnarray}
\begin{eqnarray}
\label{11}
\frac{R_{11}}{A}&+& \alpha f_R\left\{ \frac{R_{11}}{A}-\frac{R}{2}\left(\frac{n}{n+1}\right) -\frac{n}{2A}\left[ \frac{3R''}{R}+3n\frac{R'^2}{R^2}-\right.\right. \nonumber \\
3\frac{R'^2}{R^2}&+&\left.\left.\frac{R'}{R}\left(\frac{B'}{2B}-\frac{3A'}{2A}+\frac{2}{r}\right)\right]\right\} = 4\pi \left(\rho-P\right)+\Lambda \;,
\end{eqnarray}
and
\begin{eqnarray}
\label{22}
\frac{R_{22}}{r^2}&+& \alpha f_R\left\{ \frac{R_{22}}{r^2}-\frac{R}{2}\left(\frac{n}{n+1}\right) -\frac{n}{2A}\left[ \frac{R''}{R}+n\frac{R'^2}{R^2}-\right.\right. \nonumber \\
\frac{R'^2}{R^2}&+&\left.\left.\frac{R'}{R}\left(\frac{B'}{2B}-\frac{A'}{2A}+\frac{4}{r}\right)\right]\right\} = 4\pi \left(\rho-P\right)+\Lambda \;,
\label{33}
\end{eqnarray}
where the prime denotes differentiation with respect to $r$. The
fourth field equation is identical to Eq.~(\ref{33}) because of the
symmetry of the spacetime. Terms with a factor $f_R$ preceding them
are already first order in the small parameter $\alpha$ so all such
terms should be evaluated at order $\mathcal{O}(\alpha^0)$, where for example
\begin{equation}
R^{(0)} = 8\pi \left(\rho - 3P\right)+4\Lambda \\
\end{equation}
and
\begin{equation}
M^{(0)} = 4\pi \int \rho \,r^2\;dr\;.
\end{equation}
In order to motivate the form of the metric element $A(r)$ that we will be using,
we first examine the solution exterior to the star.

\subsection{The Exterior Metric}
\label{EXsec}

To solve for the exterior solution to Eq.~(\ref{EOM2}), we require
that outside the star $T_{\mu \nu} = 0$. Therefore, at
$\mathcal{O}(\alpha^0)$, the exterior metric satisfies
\begin{equation}
\label{VAC0}
R^{(0)}_{\mu \nu} = \Lambda g^{(0)}_{\mu \nu}\;,
\end{equation}
where $R^{(0)}_{\mu \nu}$ is the Ricci tensor derived from the metric
to $\mathcal{O}(\alpha^0)$. Consequently the Ricci scalar at
$\mathcal{O}(\alpha^0)$ is $R^{(0)} = 4\Lambda$. 

Note from equations (\ref{00}),(\ref{11}),
and (\ref{22}) that the $\mathcal{O}(\alpha)$ correction is multiplied by a term $f_R \propto \left[(n+1)R^{(0)}\right]^n$. For $n \ge 1$ such a theory will allow a solution with a Minkowski exterior as well as solutions with $\Lambda \ne 0$, while for $n \le -2$ the appearance of $R^{(0)}$ in the denominator requires that only solutions with $\Lambda \ne 0$ exist.

In order to calculate the corrections to the vacuum solution at
successively increasing orders in $\alpha$, we first investigate the
perturbative term in the field equation~(\ref{EOM2}), when the Ricci
curvature is constant.  At $\mathcal{O}(\alpha)$ the correction term
is proportional to
\begin{equation}
f^{(0)}_R R^{(0)}_{\mu \nu} - \frac{1}{2}g^{(0)}_{\mu \nu}\left(f^{(0)}_R R^{(0)}-f^{(0)}\right)\propto
\left(n-1 \right)R^{(0)}_{\mu \nu}\;,
\end{equation}
where we evaluated everything explicitly in terms of $R^{(0)}$ and
$R^{(0)}_{\mu \nu}$. This last relation shows that, in $f(R)$ theories
with $n=1$, the correction term in the field equation vanishes and
hence the exterior solution is identical to GR~\cite{PPD08}.

We can proceed in the same manner to arbitrary orders in
$\mathcal{O}(\alpha^m)$. The result can be formally written as
\begin{equation}
\label{VACUUM}
R^{(m)}_{\mu \nu} = g^{(m)}_{\mu \nu}\mathcal{F}\left(\alpha, \alpha^2, \dots ,\alpha^m\right)\Lambda \;,
\end{equation}
where the precise form of the function $\mathcal{F}$ is determined by the choice of the 
function $f(R)$. 

The vacuum equations, therefore, choose a unique solution, the
Schwartzchild-de Sitter metric, with
\begin{equation}
\label{ASD}
A(r) = \left(1-\frac{2M}{r}-\frac{\bar{\Lambda}r^2}{3}\right)^{-1}
\end{equation}
and $A(r)B(r) = 1$~\cite{J97}. The only difference from the general relativistic
exterior metric will be in the value of the 
effective cosmological constant, which in the case of $f(R)$ gravity 
is
\begin{equation}
\bar{\Lambda} = \mathcal{F}\left(\alpha, \alpha^2, \dots ,\alpha^m\right)\Lambda\;.
\end{equation}
As a result, the PPN parameters~\cite{W06} for an arbitrary choice of $f(R)$
will be practically those of General Relativity (see also discussion in~\cite{PPD08}).

\subsection{Interior Solution}

In the following, we shall suppress the explicit appearance of $\Lambda$ in
the field equations by the useful redefinitions
\begin{eqnarray}
\label{DEF}
&&\rho \rightarrow \rho + \Lambda \nonumber \\
&&P \rightarrow P - \Lambda \nonumber \\ 
&&M \rightarrow M +    \frac{4\pi}{3}\Lambda \, r^3 \;.
\end{eqnarray}
Subject to these normalizations and given the form of the exterior
solution we shall define
\begin{equation}
\label{ADEF}
A(r) \equiv \left[1-\frac{2M(r)}{r} \right]^{-1}\;.
\end{equation}
We will use this definition to all orders in the small parameter
$\alpha$, with the term $M(r)$ acquiring corrections at each
successive order, as it is shorthand for a metric element.

From the form of the elements of the Ricci tensor, and the above definition we obtain
\begin{equation}
\label{MDEF}
\frac{R_{00}}{2B}+\frac{R_{11}}{2A}+\frac{R_{22}}{r^2} = \frac{2M'}{r^2}\;.
\end{equation}
Combining this with equations~(\ref{00})--(\ref{22}) and evaluating the result to order $\mathcal{O}(\alpha)$ we
derive the equation for mass conservation in $f(R)$ gravity with
perturbative constraints 
\begin{eqnarray}
\label{MCON}
\frac{dM}{dr}&=& 4\pi \rho r^2 -\alpha f_R r^2\left\{ 4\pi \rho -\frac{R}{4}\left(\frac{n}{n+1}\right)-\right. \nonumber \\
\frac{n}{2A}\left[\frac{R''}{R}\right.&+&\left.\left.(n-1)\frac{R'^2}{R^2}+ \frac{R'}{R}\left(\frac{2}{r}-\frac{A'}{2A}\right)\right]\right\}.
\end{eqnarray}
The conservation
equation $\nabla^{\mu}T_{\mu \nu}= 0$ gives
\begin{equation}
\label{DT}
\frac{B'}{B} = -\frac{2P'}{\rho+P}\;,
\end{equation}
which we use in the expression for $R_{22}$ to get
\begin{equation}
\label{R22}
\frac{R_{22}}{r^2} = \frac{1}{r^2}\left[\frac{dM}{dr} + \frac{M}{r}+\frac{r}{A}\left(\frac{P'}{\rho +P}\right)\right]\;
\end{equation}
and arrive at the equation of hydrostatic equilibrium via equation (\ref{22})
\begin{eqnarray}
\label{HYDRO}
\frac{dP}{dr} &=& -\frac{A}{r^2}\left(\rho + P\right)\left\{M+4\pi P r^3 - \alpha f_Rr^3\left[ \frac{R}{4}\left(\frac{n}{n+1}\right) \right.\right. \nonumber \\
&&\left.\left. + \frac{n}{2A}\frac{R'}{R}\left(\frac{2}{r}+\frac{B'}{2B}\right) +4\pi P \right]\right\}\;.
\end{eqnarray}

Note that in solving
Eqs.~(\ref{MCON}) and (\ref{HYDRO}) in practice the evolution of the density and pressure are determined in terms of the familiar Tolman-Oppenheimer-Volkov equations
\begin{equation}
\label{dMdr}
\frac{dM^{(0)}}{dr}= 4\pi \rho_0 r^2
\end{equation}
and
\begin{equation}
\label{TOV}
\frac{dP_0}{dr}= -\frac{A^{(0)}}{r^2}\left(\rho_0+P_0\right)\left(M^{(0)}+4\pi P_0r^3\right)\;.
\end{equation}
Here $\rho_0$ and $P_0$ are understood to be the pressure and density evaluated at $\mathcal{O}(\alpha^0)$, whereas we will denote the pressure evolved via equation (\ref{HYDRO}) to $\mathcal{O}(\alpha)$ by $P_1$.

\section{Numerical Models of Neutron Stars}
\label{MRsec}

The equations we have derived so far are general and accommodate any
choice for the correction $f(R)$ to the Einstein-Hilbert action.
However, in constructing numerical models of neutron stars in $f(R)$
theories, we need to specify at this point the particular value of the
parameter $n$ we will use.

In order to address concerns for the structure and stability of
neutron stars in cosmologically motivated modifications of gravity
(see~\cite{KM08} and~\cite{BSM08}), we might consider the case
$n=-2$ (i.e., $f(R)=R^{-1}$). Since the matter density and pressure 
directly determine the Ricci scalar, we would anticipate such a term 
to be the leading order correction for small-curvature scales.
Unlike theories with additional degrees of freedom, however, and as we would
expect given the magnitude of $R$, the low-curvature 
corrections lead to no observable differences in the
structure of compact objects. Our analysis of stars with these
low-curvature corrections demand that the perturbative parameter
$\alpha$ not be significantly larger than $\Lambda$.  Such a small
correction leads to no discernible distinction from the predictions of
general relativity. 

\begin{figure}[t]
\includegraphics[angle=-90,scale=0.45]{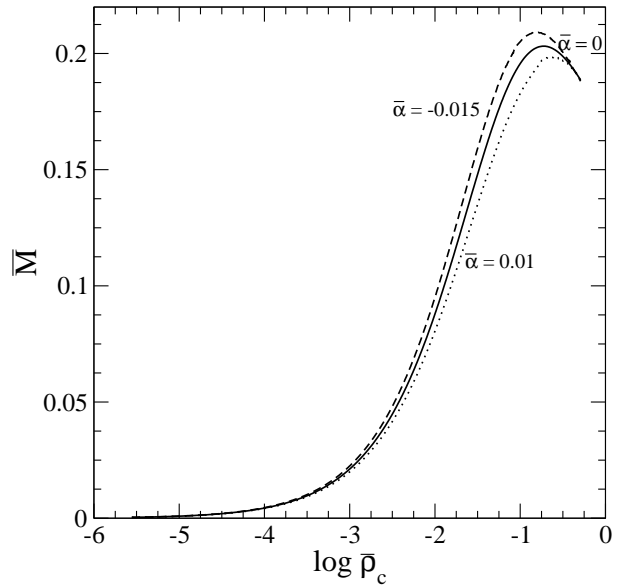}
\caption{The mass of a neutron star as a function of its central density
in an $f(R)= R^2$ gravity theory for different values of the 
small parameter $\bar{\alpha}$. The index of the polytropic
equation of state was set to $\Gamma = 9/5$.}
\label{Mrhograph}
\end{figure}

For this reason, we will study below the case with
$n=1$, i.e., gravity theories with $f(R)=R^2$.  This represents the
next to leading order correction in a high-curvature expansion of the
action. It is this regime where we expect the correction to be most
noticeable in the case of compact objects.

We choose the polytropic equation of state 
\begin{equation}
\rho = \left(\frac{P}{K}\right)^{\frac{1}{\Gamma}}+ \frac{P}{\Gamma-1}
\end{equation}
for the interior of the neutron star, where $\Gamma$ is the polytropic
index. Realistic neutron star equations of state can be parameterized
by piecewise polytropic equations of state~\cite{RLO09,OP09} with $\Gamma \simeq 1 - 3$.
 The lower the polytropic index, the stiffer the associated mass-radius relationship. For this study we have chosen $\Gamma = 9/5$ which is consistent with the constraints on $\Gamma_2$ in Ref.~\cite{OP09}.

We utilize the same dimensionless variables as in Ref.~\cite{CST94},
namely
\begin{eqnarray}
\bar{r} &\equiv & K^{-0.5/(\Gamma-1)}\;r\\
\bar{M} &\equiv & K^{-0.5/(\Gamma-1)}\;M\\
\bar{\rho} &\equiv & K^{1/(\Gamma-1)}\;\rho\\
\bar{P} &\equiv & K^{1/(\Gamma-1)}\;P\\
\bar{\alpha} &\equiv & K^{1/(\Gamma-1)}\;\alpha \;.
\end{eqnarray}
Because of this normalization of the various physical quantities, our
results are independent of the normalization $K$ of the polytropic
equation of state.  

We use a fourth order Runge-Kutte integrator with adaptive
stepsize to solve for the mass $M$ and radius $R$ of the star. We
start at the center of the star by specifying its density (and
corresponding pressure) there and integrate out to its surface defined
where the pressure vanishes.

Figure~\ref{Mrhograph} shows the dependence of the mass of a neutron
star on its central density in an $f(R)=R^2$ theory, for different
values of the small parameter $\bar{\alpha}$. The central line corresponds to
neutron stars in general relativity. As expected, for stable neutron stars the deviation from
the general relativistic case becomes significant as the central density
of the neutron star increases, since it is the matter density that
directly determines the value of the Ricci scalar curvature. Moreover,
the sign of the deviation is determined by the sign of the perturbative
parameter $\bar{\alpha}$. By properly choosing the sign and magnitude of 
this parameter, we can cause an increase or a decrease in the maximum
mass of stable neutron stars for a particular central density.
We can also support stars of a certain mass and radius for a range of central
densities and $\bar{\alpha}$.

\begin{figure}[t]
\includegraphics[angle=-90,scale=0.45]{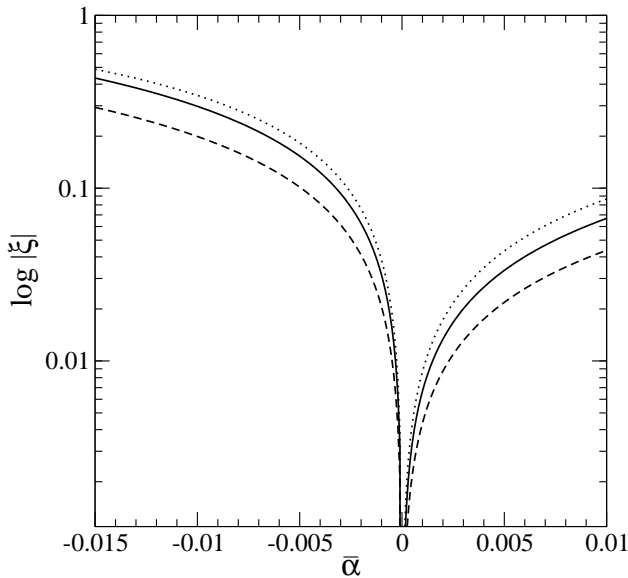}
\caption{The ratio $\xi$ (see Eq.~\ref{eq:xi}) as a function of
the parameter $\bar{\alpha}$, for stars with central densities 
$\log \bar{\rho}_c =\; -2,\;-3,\;-4$ (dotted, solid and dashed lines respectively). 
The ratio $\xi$ measures the
degree of perturbative validity of the stellar models.
A necessary condition for perturbative validity is $\xi < 1$.}
\label{stabAgraph}
\end{figure}

The maximum allowed magnitude of the deviations from the general
relativistic predictions is, of course, constrained by the requirement
that the solutions retain their perturbative validity. Though this constraint 
does not have a ready analytic expression, we can nevertheless explore
after the fact the perturbative validity of each stellar model.

In particular, as a measure of the deviations from the general relativistic solution
we choose the ratio
\begin{equation}
\xi \equiv   \left[ \frac{\bar{P}_1'}{\bar{P}_0'} \right]-1.
\label{eq:xi}
\end{equation}
This ratio varies with radius inside the neutron star. It achieves,
however, its highest value at or near the center of the star, where the density (and
hence the curvature) is large. Because we require the entire
solution to be perturbatively close to the general relativistic one,
we will evaluate the ratio $\xi$ at its maximum.
A necessary condition for perturbative validity is $\xi < 1$.

Figure~\ref{stabAgraph} shows the maximum ratio $\xi$ as a function of the parameter $\bar{\alpha}$. 
This figure demonstrates that neutron stars in $f(R)=R^2$ 
theories can certainly be treated perturbatively as long as $-0.015<\bar{\alpha}<0.01$.

\begin{figure}[t]
\includegraphics[angle=-90,scale=0.45]{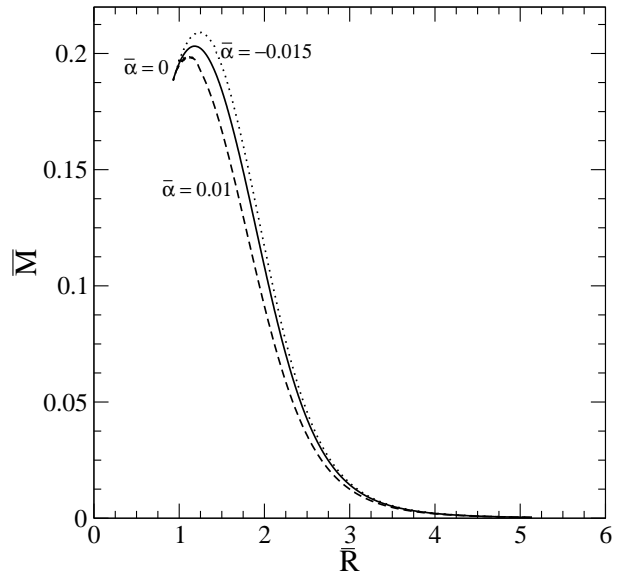}
\caption{The mass-radius relation of neutron stars for a polytropic 
equation of state with $\Gamma=9/5$, in an $f(R)=\alpha R^2$
gravity for different values of the parameter $\bar{\alpha}$.}
\label{MRgraph}
\end{figure}

Of particular interest from an observational point of view is the
mass-radius relation for neutron stars. We show this relation, for the
same polytropic equation of state, in Figure~\ref{MRgraph}. Depending
on the value and sign of the parameter $\alpha$, we obtain stars
with larger or smaller radii compared to their general relativistic
counterparts of the same gravitational mass. The extent of this variation 
is constrained by perturbative validity, which prevents the onset of dynamical
features such as spontaneous scalarization~\cite{DE92}.

\section{Discussion}
\label{DIS}

The predicted mass-radius relation for neutron stars in $f(R)$ gravity
shown above differs from that computed within general
relativity. However, very similar deviations in the mass-radius
relation can also be obtained within general relativity by simply
changing the polytropic index of the equation of state (see~\cite{OP09} for examples). Because the
equation of state of neutron-star matter is weakly constrained by
current experiments, neutron-star observables that depend only on the
mass and radius of the star cannot distinguish between small
differences in the equation of state versus small modifications to
gravity.

In~\cite{P08-2} it was shown that observables that depend also on the 
effective surface gravity of neutron stars can break, in principle, this degeneracy. In particular
it was shown that the Eddington luminosity
$L_E^{\infty}$ of a bursting neutron star depends directly on its
effective surface gravity as
\begin{equation}
L_{\rm E}^\infty \equiv
\frac{4\pi m_{\rm p} r_{\rm s}}
   {(1+X)\sigma_{\rm T}} 
   \left[\frac{z_{\rm s}(z_{\rm s}+2)}{(1+z_{\rm s})^3}\right] 
   \eta\label{eq:obs2}\;.
\end{equation}
In this equation, $m_{\rm p}$ is the mass of the proton, $X$ is the 
hydrogen mass fraction in the neutron-star atmosphere, $\sigma_{\rm T}$ is
the Thomson scattering cross section, and
\begin{equation}
z_{\rm s} = \left(1-\frac{2M}{R}\right)^{-1} -1 
\end{equation}
is the gravitational redshift from the neutron star surface. The
parameter $\eta$ is the ratio of the effective surface gravity of the
neutron star to that calculated in GR, i.e.,
\begin{equation}
\eta\equiv\frac{g_{\rm eff}}{g_{\rm GR}} 
\end{equation}
with
\begin{equation}
g_{\rm eff}\equiv \left. \frac{1}{2\sqrt{A}}\frac{d \ln B}{dr}\right|_{r=R} 
\end{equation}
and
\begin{equation}
g_{\rm GR}=\frac{1}{2R}\left[\frac{z_s\left(z_s+2\right)}{z_s+1}\right]\;.
\end{equation}

\begin{figure}[t]
\includegraphics[angle=-90,scale=0.4]{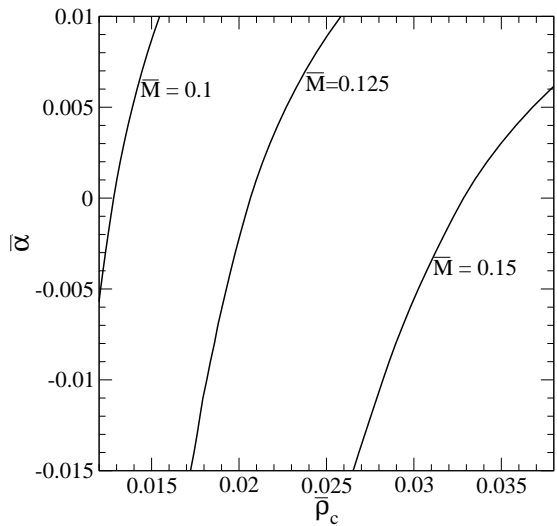}
\caption{The central density $\bar{\rho}$ of neutron stars with different
masses $\bar{M}$ as a function of the parameter $\bar{\alpha}$.
Larger positive deviations from general relativity require larger
central densities for the same neutron-star mass and, therefore,
lead to shorter cooling times. On the other hand, larger negative
deviations require smaller central densities and lead to longer
cooling time.}
\label{alpharhocgraph}
\end{figure}

We can calculate easily the value of the parameter $\eta$ for the 
$f(R) = R^2$ theory considered here.
From the conservation equation~(\ref{DT}) we can write
\begin{equation}
\label{geff}
g_{eff}=-\frac{1}{\sqrt{A}}\frac{P'}{\left(\rho+P\right)}\;.
\end{equation}
We can then evaluate the hydrostatic equilibrium equation~(\ref{HYDRO}) 
to first order in $\alpha$ by noting that
\begin{equation}
\frac{R^{(0)'}}{A^{(0)}} = -8\pi\left(\frac{\partial \rho_0}{\partial P_0}-3\right)\frac{\left(\rho_0 + P_0\right)}{r^2}\left(M^{(0)}+4\pi P_0 r^3\right)\;.
\end{equation}
As a result equation~(\ref{geff}) becomes
\begin{eqnarray}
g_{eff} &=& \frac{ \sqrt{A^{(1)}} }{r^2} \left(M^{(1)}+4\pi P_1 r^3\right) -\alpha \left\{8\pi \left(\rho_0+P_0\right)\sqrt{A^{(0)}} r \right.  \nonumber \\
 && \left[ 2\pi \left(\rho_0-3P_0\right) + \frac{2}{r^3}\left(3-\frac{\partial \rho_0}{\partial P_0}\right)\left(M^{(0)}+4\pi r^3 P_0\right) \right. \nonumber \\
&& \left. \left. \left.  +\frac{A^{(0)}}{r^4}\left(3-\frac{\partial \rho_0}{\partial P_0}\right) \left(M^{(0)} +4\pi r^3 P_0\right)^2 \right]\right\}\right|_{r=R} 
\end{eqnarray}
At the surface layer of the neutron star $\rho = P = 0$ and hence
\begin{equation}
g_{eff} = \frac{\sqrt{A^{(1)}}}{R^2}M^{(1)}\;.
\end{equation}
Which has the same dependence on mass and radius as $g_{GR}$ does.
As a result measuring $\eta$ alone will not suffice to break the 
degeneracy due to the equation of state.

Nevertheless constraining observationally the cooling rates of neutron stars can offer a discriminant. 
A neutron star cools
both through photon and neutrino emission. The photon luminosity is determined by the 
temperature at the photosphere, which in turn depends on the density of the photosphere.
 However neutrino cooling, which depends more sensitively on temperature than photon
  cooling does, becomes dominant for neutron stars with temperatures above $10^{10}K$, 
  and indeed is the primary mechanism of cooling for young neutron stars 
 (see~\cite{PGW06} for a detailed review). The high temperature and low interaction rate
 make neutrino cooling particularly sensitive to the central density of the neutron star.
Figure~\ref{alpharhocgraph} shows the relation between the parameter $\bar{\alpha}$ and
the central density of a neutron star, for three different values of the mass
$\bar{M}=0.15$, $0.125$, and $0.1$. Large positive deviations from general
relativity, as measured by the parameter $\bar{\alpha}$ require larger
central densities for neutron stars of a given mass, whereas the opposite
is true for large negative deviations. As a result, because the cooling timescale
scales with central density, observations of the surface temperatures of young
neutron star can lead to useful constraints on the deviations from general
relativity in an $f(R)$ gravity model, especially if the neutron-star masses are
known.

We will study the constraints imposed on $f(R)$ gravity by current
measurements of cooling rates of neutron stars in our
galaxy in a forthcoming paper.

\acknowledgements

We wish to thank the authors of~\cite{ADE10} for important and helpful correspondence.

\end{document}